\documentclass[sigplan,10pt]{acmart}

\usepackage{booktabs}   %% For formal tables:
\usepackage{subcaption} %% For complex figures with subfigures/subcaptions
\makeatletter\if@ACM@journal\makeatother
\acmJournal{PACMPL}
\acmVolume{1}
\acmNumber{1}
\acmArticle{1}
\acmYear{2017}
\acmMonth{1}
\acmDOI{10.1145/nnnnnnn.nnnnnnn}
\startPage{1}
\else\makeatother
\acmConference[PLATEAU'17]{PLATEAU'17 Workshop on Evaluation and Usability of Programming Languages and Tools}{October 23, 2017}{Vancouver, CA}
\acmYear{2017}
\acmISBN{978-x-xxxx-xxxx-x/YY/MM}
\acmDOI{10.1145/nnnnnnn.nnnnnnn}
\startPage{1}
\fi

\setcopyright{none}             %% For review submission
\usepackage[T1]{fontenc}
\usepackage[utf8]{inputenc}

\usepackage{pifont}

\def\Check{\ding{51}}
\def\NoCheck{\ding{55}}

\usepackage{tabularx}
\newcolumntype{C}{>{\centering\arraybackslash}X}
\newcolumntype{L}{>{\raggedright\arraybackslash}X}
\newcolumntype{R}{>{\raggedleft\arraybackslash}X}
\newcommand*\rot{\rotatebox[origin=B]{90}}
\usepackage{booktabs}

\usepackage[nohyperlinks,nolist]{acronym} % options: nolist printonlyused

\begin{document}

%% Title information
\title{Development of a Web Platform for Code Peer-Testing}      
%\title[Peer-Testing]{Development of a Web Platform for Code Peer-Testing}         %% [Short Title] is optional;
                                        %% when present, will be used in
                                        %% header instead of Full Title.
%\titlenote{with title note}             %% \titlenote is optional;
                                        %% can be repeated if necessary;
                                        %% contents suppressed with 'anonymous'
%\subtitle{Subtitle}                     %% \subtitle is optional
%\subtitlenote{with subtitle note}       %% \subtitlenote is optional;
                                        %% can be repeated if necessary;
                                        %% contents suppressed with 'anonymous'

%% Author information
%% Contents and number of authors suppressed with 'anonymous'.
%% Each author should be introduced by \author, followed by
%% \authornote (optional), \orcid (optional), \affiliation, and
%% \email.
%% An author may have multiple affiliations and/or emails; repeat the
%% appropriate command.
%% Many elements are not rendered, but should be provided for metadata
%% extraction tools.

%% Author with single affiliation.
\author{Manuel Maarek}
%\authornote{with author1 note}          %% \authornote is optional;
                                        %% can be repeated if necessary
\orcid{0000-0001-6233-6341} % Maarek             %% \orcid is optional
\affiliation{
%  \position{Assistant Professor}
  \department{School of MACS}              %% \department is recommended
  \institution{Heriot-Watt University}            %% \institution is required
%  \streetaddress{Riccarton}
  \city{Edinburgh}
  \state{Scotland}
%  \postcode{EH14 4AS}
  \country{UK}
}
\email{M.Maarek@hw.ac.uk}          %% \email is recommended

%% Author with two affiliations and emails.
\author{Léon McGregor}
%\authornote{with author2 note}          %% \authornote is optional;
                                        %% can be repeated if necessary
%\orcid{nnnn-nnnn-nnnn-nnnn}             %% \orcid is optional
\affiliation{
%  \position{Student}
  \department{School of MACS}              %% \department is recommended
  \institution{Heriot-Watt University}            %% \institution is required
%  \streetaddress{Riccarton}
  \city{Edinburgh}
  \state{Scotland}
%  \postcode{EH14 4AS}
  \country{UK}
}
%\email{first2.last2@inst2a.com}         %% \email is recommended

%% Paper note
%% The \thanks command may be used to create a "paper note" ---
%% similar to a title note or an author note, but not explicitly
%% associated with a particular element.  It will appear immediately
%% above the permission/copyright statement.
%\thanks{QAA Heriot-Watt}                %% \thanks is optional
                                        %% can be repeated if necesary
                                        %% contents suppressed with 'anonymous'

%% Abstract
%% Note: \begin{abstract}...\end{abstract} environment must come
%% before \maketitle command
\begin{abstract}
  As part of formative and summative assessments in programming
  courses, students work on developing programming artifacts following
  a given specification. These artifacts are evaluated by the
  teachers. At the end of this evaluation, the students receive
  feedback and marks. Providing feedback on programming artifacts is
  time demanding and could make feedback to arrive too late for it to be
  effective for the students' learning. We propose to combine software
  testing with peer feedback which has been praised for offering a
  timely and effective learning activity with program testing.

  In this paper we report on the development of a Web platform for
  peer feedback on programming artifacts through program testing. We
  discuss the development process of our peer-testing platform
  informed by teachers and students.
\end{abstract}

%% 2012 ACM Computing Classification System (CSS) concepts
%% Generate at 'http://dl.acm.org/ccs/ccs.cfm'.
\begin{CCSXML}
<ccs2012>
<concept>
<concept_id>10003456.10003457.10003527.10003531.10003751</concept_id>
<concept_desc>Social and professional topics~Software engineering education</concept_desc>
<concept_significance>500</concept_significance>
</concept>
<concept>
<concept_id>10003456.10003457.10003527.10003531.10003533</concept_id>
<concept_desc>Social and professional topics~Computer science education</concept_desc>
<concept_significance>300</concept_significance>
</concept>
<concept>
<concept_id>10011007.10011074.10011099.10011102.10011103</concept_id>
<concept_desc>Software and its engineering~Software testing and debugging</concept_desc>
<concept_significance>500</concept_significance>
</concept>
<concept>
<concept_id>10010405.10010489.10010492</concept_id>
<concept_desc>Applied computing~Collaborative learning</concept_desc>
<concept_significance>300</concept_significance>
</concept>
</ccs2012>
\end{CCSXML}

\ccsdesc[500]{Social and professional topics~Software engineering education}
\ccsdesc[300]{Social and professional topics~Computer science education}
\ccsdesc[500]{Software and its engineering~Software testing and debugging}
\ccsdesc[300]{Applied computing~Collaborative learning}
%% End of generated code

%% Keywords
%% comma separated list
\keywords{peer testing, peer feedback, software testing}  %% \keywords is optional

%% \maketitle
%% Note: \maketitle command must come after title commands, author
%% commands, abstract environment, Computing Classification System
%% environment and commands, and keywords command.
\maketitle

\section{Introduction}
\label{sec:introduction}

Students in \ac{CS} courses take part in individual coursework
exercises in the form of programming tasks. Such programming
coursework often play an important role in their assessment and are an
opportunity to train and develop the students programming skills, as
such the feedback they receive on the artifacts they develop is
important for their learning. While students receive face-to-face
feedback on their programs during computer lab sessions, the bulk of
feedback accompanies the grading of the artifacts they have submitted
for assessment. As producing programming feedback is a time consuming
task, a number of experiments have been conducted to use testing of
programming artifacts to automatically produce feedback to students
where failed and passed tests construct the program feedback. In
\ac{CS} and \ac{HE} in general, peer feedback which involves students
providing feedback on their peers' work is often praised for offering
a timely and effective learning activity.

We propose to combine peer feedback and software testing to offer
timely feedback on programming artifacts. In this approach, after
submitting their own programming artifacts, students would engage in a
peer feedback activity on their peers' submissions. The main medium
for feedback is software tests that students would run on their peers'
submissions.  To facilitate this peer feedback activity, we have
developed a peer-testing Web platform which manages the submission of
solutions and tests, their execution, and the sharing of results and
textual feedback.  Ideally, the peer feedback that takes place within
the Web platform will provide more immediate and useful feedback on
coursework than having to wait for a teacher, and the involvement in
the peer-testing activity invites students to engage in critical
thinking about the programming process.

This paper reports on the development of our peer-testing platform
which was informed by teachers and students.
The paper:
\begin{itemize}
\item Presents peer-testing as peer feedback on programming
  artifacts through program testing.
\item Discusses aspects and features teachers expected from a peer
  feedback system for code peer-testing.
\item Report students opinions on the Web platform after experiencing
  peer-testing.
\item Describes the peer-testing platform we have implemented and the
  stages making a peer-testing activity.
\end{itemize}

\paragraph{Plan}

In Section~\ref{sec:background} we give the background of this work on
peer feedback, peer assessment and software testing. We describe in
Section~\ref{sec:requirements} how we derived the key requirements for
the peer-testing platform from discussions with teachers and students,
and we give an overview of the platform we have implemented in
Section~\ref{sec:platform}.  We then discuss related works in
Section~\ref{sec:related-works} before drawing perspectives for this
work in Section~\ref{sec:conclusion}.

The work presented here took place at Heriot-Watt University during
Léon McGregor's BSc Honours project~\cite{McGregor_BSc-2017} and as
part of the University's Learning and Teaching Enhancement project
lead by Manuel Maarek. Aspects of the project, in particular its
impact on students' transition from passive learners to critical
evaluators was presented at Horizons in STEM Higher Education 2017
Conference~\cite{GroHamKumMaaMcGrShaWelZan_STEM-HE-2017}.

\section{Background}
\label{sec:background}

\subsection{Peer Feedback and Peer Assessment}
\label{sec:background-peer-feedback}

Peer assessment is a process in which students assess each other. It is in
contrast to the more traditional stance where a teacher performs the
assessment. As defined by \citet{topping_peer_2009},
\begin{quote}
  ``Peer assessment is an arrangement for learners to consider and
  specify the level, value, or quality of a product or performance of
  other equal-status learners.''
\end{quote}
That is to say, students with a similar level of education assessing
the work of each other to give critical feedback and discussion. This
could be done in many ways, such as between pairs or in groups, and
can be performed on any number of different activities from
programming exercises to oral reports.

%\paragraph{Why use it?}

Peer assessment is a process with many benefits to participants in
education. \citet{sadler_impact_2006} have suggested the
following concepts that peer assessment can help with
\begin{itemize}
\item Peer assessment is more immediate, so students can get more
  feedback, and sooner;
\item Students performing marking can reduce the workload for
  teachers;
\item The process of checking and thinking about another students
  answer can improve a students own understanding;
\item Peer assessment can help students better understand testing and
  can become aware of their own strengths and weaknesses;
\item Following peer assessment students can gain an improved attitude
  towards the process of learning as a whole.
\end{itemize}

Peer assessment can offer much help towards education of students, but
it would be worthwhile to know just which aspects are the most
useful. A study conducted by \citet{li_assessor_2010} investigated the
peer assessment process with the aim of discovering which part of it
is most useful to the students involved: being an assessor or being
assessed. To study this, undergraduate student teachers were given the
task of creating a WebQuest project (activities for student to learn
from Internet resources) This was then marked by independent
assessors, and the student teachers were given a chance to provide
feedback on other student teachers' WebQuests. Following this, the
feedback was returned and student teachers were given another chance
to improve their project, and it was marked again. The quality of the
peer feedback itself was also checked by the independent markers. The
study found that

\begin{quote}
  ``there was a significant relationship between the quality of the
  peer feed-back the students provided for others and the quality of
  the students' own final projects''
\end{quote}

The findings of the investigation would suggest that the actual
exercise of providing feedback to others (acting as an assessor) is a
worthwhile process for learning from. This study also concluded that
there was no reasonable link between the feedback itself as a learning
tool, suggesting that the act of giving feedback itself is more
valuable and that low quality feedback does not harm the learning
experience.

With the knowledge that peer assessment can be useful, it is important
to know how a peer assessment should be conducted. A study performed
in a classroom environment by \citet{SmiTesKraLin_ICER-2012}
focused more on the use of peer assessment as a tool for teaching
testing of code, in addition to the existing course.
Over the course of this study, which took place using coursework from
a 12-week university course, the following was completed for each
coursework: submitting solutions, then submitting peer reviews (which
includes a description of the testing that they performed on another
solution, and the results of this), and then a review of the
peer assessment (including what was learnt, an evaluation of feedback
on their own solution, and optionally a corrected solution).
One particularly noteworthy aspect of this use of peer assessment was
the double-blind nature, ensuring anonymity. Students would not be
aware of who they were marking, or were marked by. To enforce this
completely, submitted code was obfuscated (Java sources into Byte
code). One advantage of this is that it strips out identifying
variable names and comments, which could identify other
students. However, a downside of using byte code is that it can make
it difficult to do in-depth analysis of the source structure which may
make it harder to write complete test cases.
The study identified two key features that assignments for
peer assessment need to have: a well-defined interfaces, and freedom
for implementation.  In addition to this,
\citet{SmiTesKraLin_ICER-2012} has found that it was possible to
integrate the peer assessment process without having to significantly
alter the existing course material, and the students taking part
enjoyed the experience. This shows promise, as it could indicate many
\ac{CS} courses (that offer coursework meeting the requirements),
could be modified to include their own peer assessment exercises.

Peer assessment can prove to be a very valuable experience for
students. \citet{falchikov_improving_2013} has collected various case
studies of past peer assessments, and the following aspects can be
found:
\begin{itemize}
\item If the marking criteria are properly explained, there is often
  no significant difference between the marks awarded by students and
  those that would be awarded by teachers. This would tend to indicate
  that students do assess each other fairly.
\item One of the most important aspects of peer assessment is the
  ability of the student to learn how to assess other students and
  from this learn how to critically assess and improve their own work.
\item It is important to make sure students feel confident, otherwise
  they may not assess their peers as honestly as they might otherwise
  have done. Some students will feel conflicted about marking their
  peers, particularly if they might have to give low marks.
\item During peer assessment more benefits may come from students
  assessing multiple solutions, rather than each focusing on one.
\end{itemize}

\subsection{Software Testing Methods}
\label{sec:background-testing}

When considering the testing of software, there are several ways that
this can be done. To produce a Web platform that enables peer assessment,
an appropriate testing methodology will need to be selected. Based on
suggested testing styles from  \citet{laboon_friendly_2016},
some testing methodologies were considered for inclusion.

\paragraph{Linting} Performing basic checks for 'smells' of bad code
  design, such as identifying unused variables or methods. Not
  particularly useful from a perspective of checking code is correct,
  but is still useful in terms of producing good quality code.
\paragraph{Unit testing} The use of xUnit style tests. Often used in
  \ac{TDD}, these could be useful in detecting flaws if
  written post-development by a peer assessor. This would need the
  assessor to be familiar with how to write xUnit style tests, and how
  they should be structured, or it would not be viable. (The
  prototypes and current implementation presented in this paper use
  the unit testing form of testing)
\paragraph{Expected output testing} Running a program with some input and
  comparing the output to what is expected. The issue here is that the
  peer assessor first needs to know what the correct output is. Unlike
  unit testing, this would simply involve a series of inputs, and a
  series of expected outputs, and the task of checking the correctness
  from these tests is left to the website.
\paragraph{Scenario Testing} Simulate an actual usage scenario. Assessors
  would use solutions as if they were third party libraries (a
  \emph{black box}), and develop their own programs that would use
  these as if in an actual use case. These programs would perform
  checks to make sure the solutions being tested were running as
  expected. This is more involved than unit or expected output
  testing, as it might be better placed to discover side effects of
  continued use of the solution.
\paragraph{Property based testing} Using a proof checker or random testing
  such as QuickCheck to ensure that a program is acting correctly. This
  would place additional overhead onto the students as assessors, as
  they would need to learn and understand the annotations used by a
  proof checker, but could provide a lot of coverage of possible
  inputs.

\section{Informed Peer-Testing Requirements}
\label{sec:requirements}

The peer-testing platform is a tool for enabling peer-testing of
program artifacts as a learning activity in the classroom. As such its
main users are the teachers and the students. We have planned the
development of the platform into steps involving teachers and students
to gathered information about their needs and perspectives on a
platform for peer-testing. At each steps we developed a prototype that
was used as a base for discussion. In the first step, we implemented an
initial prototype following a concise set of requirements (see
Section~\ref{sec:initial-requirements}). In a second step teachers
were invited to discuss these requirements and propose new ones (see
Section~\ref{sec:teacher-requirements}). A second prototype was then
developed implementing some of these requirements, and was evaluated
by a pool of students who suggested improvements (see
Section~\ref{sec:student-requirements}). The current version of the
peer-testing platform presented in Section~\ref{sec:platform}
implements some of these suggestions.

\subsection{Initial Requirements}
\label{sec:initial-requirements}

In our initial design, the peer-testing process was asynchronous where
each student would (1) submit a solution as a source code file of
their program. The system would then (2) pair the students into
tester/developer before it enters in peer-testing mode, where each
student would (3) as tester submit tests, which are run against their
peer's solution. Each student would then (4) submit a test with
textual feedback which the (5) student as developer would see
displayed. As part of the coursework tasks, each student would submit
a report discussing their involvement in peer-testing and how they
would update their program to take into account the feedback they
received or gave. This initial simple workflow was implemented and
used as a base for discussion with teachers.

\subsection{Focus Group Discussions with Teachers}
\label{sec:teacher-requirements}

To get information on the teachers' perspective on peer-testing, we
invited seven \ac{CS} lecturers from our department to discuss the
initial set of requirements and to propose features they consider to be
essential for a peer-testing platform in the classroom. We summarise
here the main topics of discussions.

\paragraph{Self Testing}

To make sure students are ready to interact with peer-testing it is
essential that they test their own implementation. So the platform
should allow testing own implementations form the start (we have
possible in Stage 1, see Section~\ref{sec:platform}). The ability to
craft tests based on someone else's implementation could also be
allowed prior to peer-testing but would require to hide the source
code of the peer's implementation (black-box testing) to avoid
plagiarism issues (we have made this possible only on the teacher's
provided solution, see Section~\ref{sec:platform}). Some own tests
could be made private so that they are not shared with peers (we added
this feature but retracted it after it created confusion with
students, see Section~\ref{sec:student-requirements}).

\paragraph{Teacher Monitoring}

For the teacher to monitor the peer-testing activity and the learning
the students go through, the peer-testing platform should offer the
ability to track the students' attempts at testing their own solution
and their peers' solutions. Such monitoring view could form a
testing-based learners log which the teacher would use in assessment.
Tracking everything a student has done and putting this in one place
could offer a way for the teacher to give feedback on a given
coursework (this feature was implemented but is not presented in this
paper).

\paragraph{Coursework Artifacts}

The coursework specification needs to provide interfaces where testing
will be conducted. These interfaces could give rise to signature case
tests that the teacher provides to ensure that the student solution
met the interface required (we have made such teacher provided tests
possible, see Section~\ref{sec:platform}).

Tests could be crafted based on input only, the expected output is
then produced by a standard solution or oracle that the teacher
provides. Students would have the ability to run their own test cases
on the oracle solution without having access to its source code (in
the current platform, testing is done with unit tests, we have made
such teacher oracle implementation part of the coursework setup, see
Section~\ref{sec:platform}).

Providing a signature test case and an oracle is to assist the student
in crafting solutions and tests. This could be complemented by
additional analysis or testing providing automated feedback to the
students, but this goes beyond the core purpose of the platform.

\paragraph{Technical Transparency}

The testing framework of the platform should not hide the technical
details from the students: the commands that are run should be
displayed; if issues are detected automatically with pre-set
verification at compile time or at runtime, the students should be
invited to alter their submission rather than having the system
pre-processing it (in the current implementation the processing of
output is limited to hiding execution paths, command lines are not
printed).

\paragraph{Anonymity}

As our platform is intended for peer feedback, we asked the question
\emph{should peer-testing be anonymous?} This would require to hide
identifications in submitted files (or to ask the students to do
so). The advantages are numerous as it extends to gender anonymity,
campus anonymity, and would allow for teacher-crafted submissions and
tests to be included as peer submissions.  An alternative to asking
the students to anonymise their submission is to sanitise the
solutions submitted to keep identities anonymous by for instance
removing source code comments, but this would hinder code comprehension.

During a peer assessment / automated testing experiment performed by
\citet{sitthiworachart_effective_2004}, the students remarked about
the use of anonymity. One students response suggested that in a
non-anonymous peer assessment exercise, their marking would be
influenced if they knew who it was they were marking. Another student
said they would decline to take part if the peer assessment was not
conducted anonymously. What can be drawn from these comments is that
anonymity is clearly very important to those taking part in the peer
assessment. But the opinions of anonymity might not correlate with the
actual effect it has.
To study the effect that anonymity has, as opposed to just opinions
about it, a study was conducted by \citet{li_role_2016}. This study
aimed to investigate just how effective anonymity is when it comes to
making peer assessment more effective, and whether any negative impact
from a lack of anonymity can be mitigated.
This quasi-experimental study was conducted with some in-training
teachers, and aimed to see which is the most effective method of /
using / while conducting a peer assessment exercise: Having assessors
and assessees know each others identities, remain anonymous, or know
identities while having received training. The training in this case
involved watching a video that described some of the stresses and
concerns that arise during peer assessment, and various forms of
discussion regarding this.
The study did not cover the case of being anonymous and getting
training. This was because the training was intended as a fallback for
when anonymity is not possible. But this does invite the question of
just how effective would anonymity be if training were offered as
well.
The study found: that ``anonymity improves student performance in peer
assessment'', that if anonymity cannot be guaranteed, negative effects
arising from this can be offset using training,and that anonymity does
not reduce the pressure and tension related with peer assessment.

Currently our platform does not perform any automated anonymisation of
the code submitted, instead the students are provided with
instructions to remove their name and any other mean of identification
from their code before submission or during the online discussions.

\subsection{Experiment, Questionnaire and Focus Group Discussion with Students}
\label{sec:student-requirements}

We conducted an experiment with students studying in year 2 and 3 of
the \ac{CS} department of Heriot-Watt University.
The participants were invited to prepare solutions for two small
programming exercises (in Python). For each exercise, each student was
paired with another student and invited to test their peer's
solution. The peer-testing was to be done via the prototype Web
platform for the first exercise and via email exchanges for the second one.
Following this experiment, the participants were asked to complete a
questionnaire and were invited to take part in a focus group
discussion in order to tease out the ways to improve the website
further and to gain an insight into their opinions on the peer-testing
activity, this latter aspect is discussed
in~\cite{GroHamKumMaaMcGrShaWelZan_STEM-HE-2017}.
As the programmes are taught in two campuses of the University, namely
Edinburgh and Dubai, we had participants from both locations. The way
students perceived positively peer-testing as an opportunity for
cross-campuses interactions is also discussed
in~\cite{GroHamKumMaaMcGrShaWelZan_STEM-HE-2017}. While the
participants were well distributed across our campuses (5 in
Edinburgh, 6 in Dubai) and in terms of gender (4 identified as female,
7 identified as male), their small number does not allow for
quantitative analysis of the feedback. We could however conduct
qualitative analysis of
the questionnaire answers, and the focus group discussion that
followed the experiment. These are discussed below.  In general, the
comments made by the participants suggest that such website would be
welcome in future iterations of \ac{CS} courses.

\paragraph{Importance of Training}

Points raised in answers to the question \emph{Was the website
  behaving correctly? Were there any bugs?}  suggested that the
participants were put off by the number of failing test cases.
Multiple respondents felt that the failing tests limited their ability
to provide meaningful feedback. Further investigation into the tests
that were submitted to the website, suggested that most of these tests
were not failing as a result of website implementation, but were
actually related to issues in how participants had written the 
testing code in python. The way to reduce this issue is to train students to
writing test cases and to better explain how the website will run the
tests.

The peer-testing platform should first be used in formative lab
sessions exercises before being used for an summative coursework to
make sure students are familiar with its use.

\paragraph{Peer Group}

The issue of students not engaging with the peer activity occurred
during the experiment: since participants were paired for each
exercises, some participants were prevented from taking part since
their paired peer did not engage with the exercise.

Organising peer groups of students rather than pairs of students
should alleviate this issue, as suggested earlier in
Section~\ref{sec:background-peer-feedback}. Additionally, peer
feedback practices recommend to put in place incentives for students to
engage in the peer-testing activity such as to allocate some marks for
taking part and for submitting a reflective report after the activity.

\paragraph{Feedback Discussion}

One issue that appeared in both the survey, and while participants
interacted with the website, was that once a feedback comment has been
submitted for a test match, it cannot be modified. This caused issues
for one participant as they had accidentally clicked the submit
button. A solution is to allow feedback to be changed after it has
been submitted, although a history of responses would need to be kept.
Investigating some of the email-based peer-testing feedback, it looked
as though many of the students were providing feedback on their peer's
solution as well as on the test cases submitted by their peer's.
In the discussion session, a participant suggested to use a 2-way
discussion on the website.

A 2-way discussion between pairs of participants (currently
implemented, see Section~\ref{sec:platform}) or a forum-style
discussion within peer groups would be beneficial in giving more
opportunities for feedback.

\paragraph{Usability}

The evaluation revealed many issues with the usability of the
website. The main design of the website and the lack of clear
explanation as to how it would function caused some confusion in both
how to use the website, and how the various features that were
available fit together. This in turn resulted in requiring a lot of
effort on the part of the users to get the functionality working
right. A number of improvements were carried out since the
prototype used in the experiment (contextual help, removing the
ability to mark some uploads as \emph{private}, re-design some
layouts, mobile-friendly layout).

\section{Peer-Testing Platform}
\label{sec:platform}

The peer-testing platform we are presenting and reporting on in this
paper was implemented in Python using the Django Web framework. A
web-page\footnote{\url{http://www.macs.hw.ac.uk/~mm894/peer-testing}}
dedicated to the peer-testing project provides additional information,
and the source code is available on
GitHub\footnote{\url{https://github.com/peergramming/peer-testing}}.
The current implementation supports Python (with the \texttt{unittest}
package for running tests) and Java (with JUnit).

In this section we describe the workflow of using peer-testing with
our platform within a coursework assignment.

\paragraph{Peer-Testing Stages}

\begin{figure*}
  \begin{minipage}{.45\linewidth}
\begin{minipage}[t]{1\linewidth}
  \centering
  \includegraphics[width=1\linewidth]{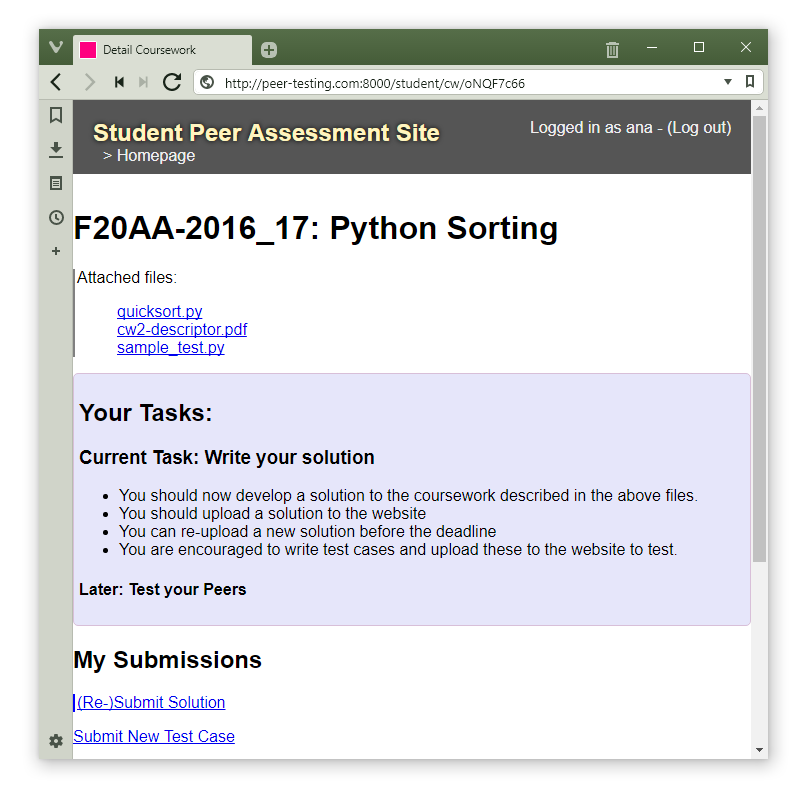}
  \subcaption{Stage 0: Coursework Setup}
  \label{fig:stage-0}
\end{minipage}%

\begin{minipage}[t]{1\linewidth}
  \centering
  \includegraphics[width=1\linewidth]{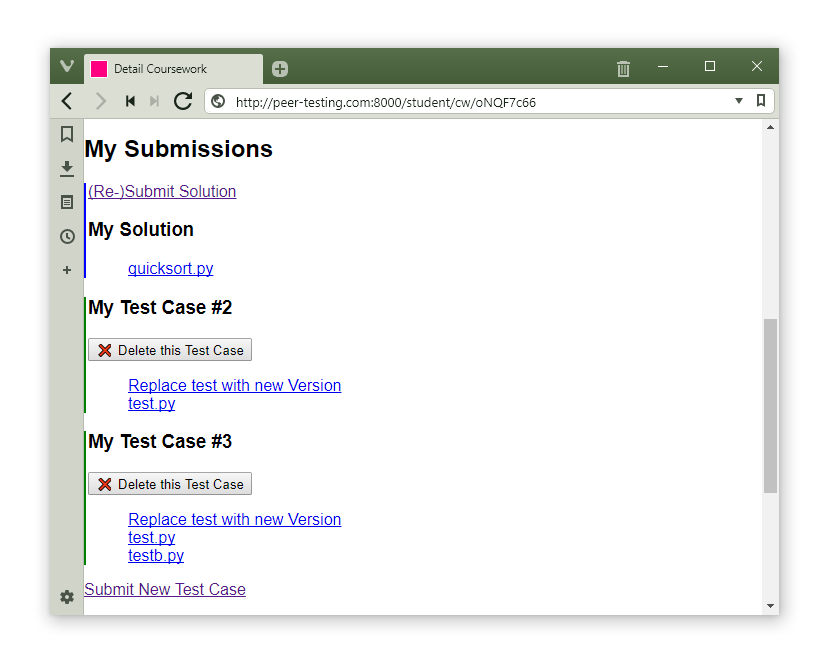}
  \subcaption{Stage 1: Development \& Self-Testing -- Uploading
    solutions and tests}
  \label{fig:stage-1-1}
\end{minipage}%

\begin{minipage}[t]{1\linewidth}
  \centering
  \includegraphics[width=1\linewidth]{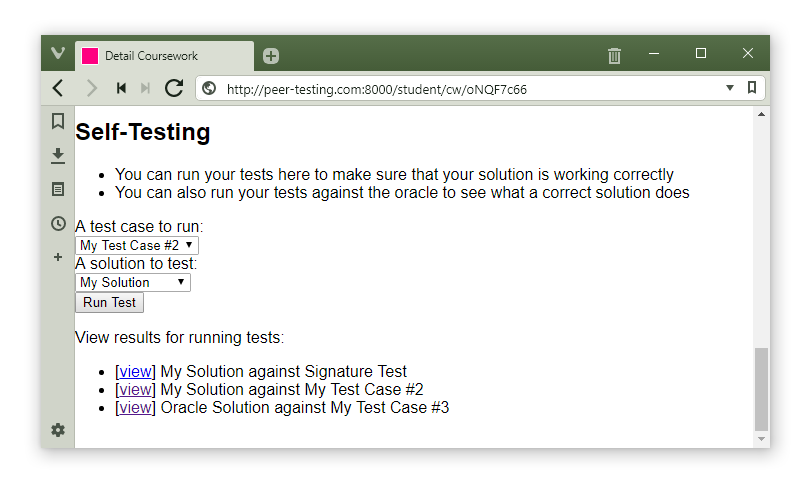}
  \subcaption{Stage 1: Development \& Self-Testing -- Running tests }
  \label{fig:stage-1-2}
\end{minipage}%
  \end{minipage}\hfil%
  \begin{minipage}{.45\linewidth}
\begin{minipage}[t]{1\linewidth}
  \centering
  \includegraphics[width=1\linewidth]{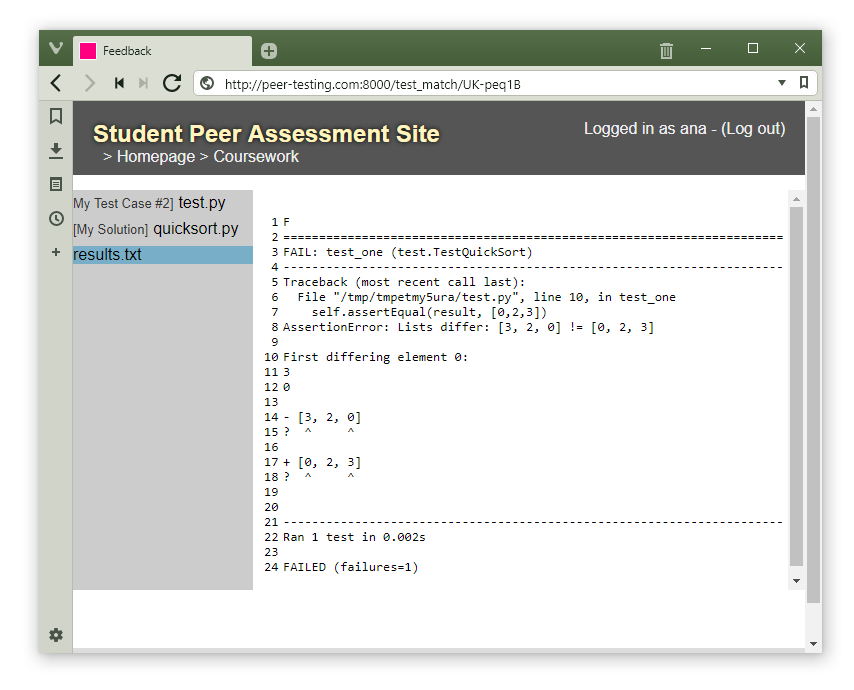}
  \subcaption{Stage 1: Development \& Self-Testing -- Viewing test results}
  \label{fig:stage-1-3}
\end{minipage}%

\begin{minipage}[b]{1\linewidth}
  \centering
  \includegraphics[width=1\linewidth]{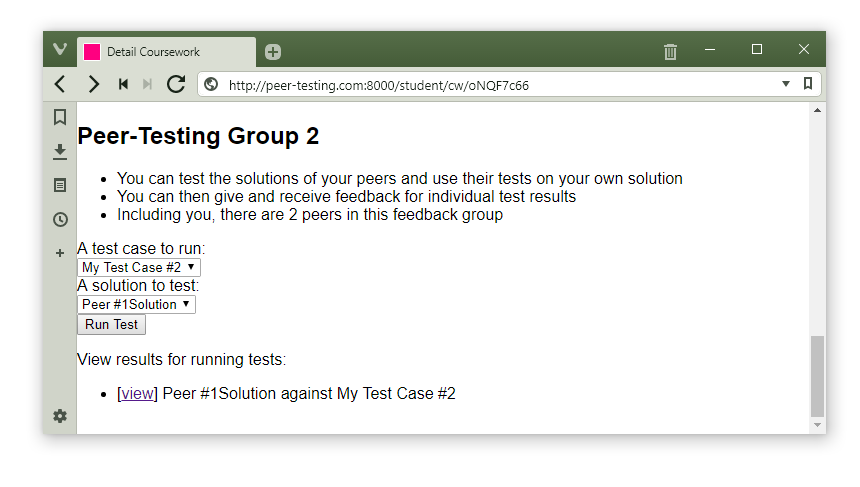}
  \subcaption{Stage 2: Peer-Testing \& Feedback -- Running tests}
  \label{fig:stage-2-1}
\end{minipage}%

\begin{minipage}[t]{1\linewidth}
  \centering
  \includegraphics[width=1\linewidth]{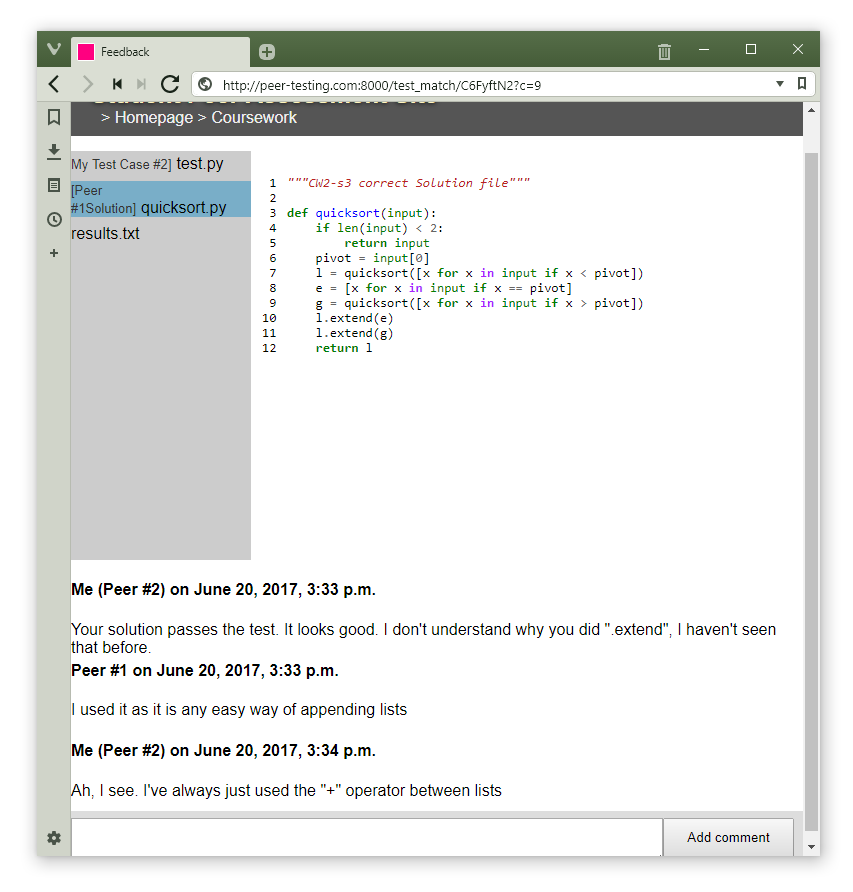}
  \subcaption{Stage 2 Peer-Testing \& Feedback -- Tests results and
    feedback discussion}
  \label{fig:stage-2-2}
\end{minipage}%
  \end{minipage}
\caption{Screenshots of Peer-Testing Stages}\label{fig:stages}
\end{figure*}

The website helps to split up the completion of the coursework into
several stages, each with different ability to interact with the
website and coursework. These stages, and interactions from the
teachers and students involved are detailed in Table~\ref{tab:stages},
and screenshots resulting form some website actions are shown in
Figure~\ref{fig:stages}.

In the following we will illustrate these stages of a peer-testing
activity with a small exercise requiring the students to implement a
sorting algorithm in Python.
\begin{description}
\item[Stage 0: Coursework Setup] The teachers prepares the materials
  for the coursework exercise. This would be a document the giving the
  exercises instructions and learning objectives, in this case
  implementing the QuickSort algorithm in Python. The teachers could
  also provide sample solutions and tests.

  A sample solution could be an implementation that meets the
  specification so that student can evaluate the effectiveness of
  their test cases, we call such sample implementation an
  oracle. Although students could run their tests on the oracle, its
  source code will not be accessible to them. Another sample solution
  could be a template that student could use as a starting point.

  A sample test could be a simple test case which is used to make sure
  student solution matches the interfaces of the specification, we
  call such test a signature test case. Other sample test cases could
  be provided that student solution are expected to pass for getting
  certain marks.

  Once this stage is achieved the students would have access to the
  coursework information through the coursework's home page, see
  Figure~\ref{fig:stage-0}.
\item[Stage 1: Development \& Self-Testing] The students are working
  individually on their implementation. They can upload the source
  files of their solutions as well as tests they have prepared, see
  Figure~\ref{fig:stage-1-1}. They can then start running their
  implementation against the provided test cases or their own tests,
  they could also run these tests on the oracle, see
  Figure~\ref{fig:stage-1-2}. Once a run has been requested and
  performed by the platform, the student could investigate its result
  as shown in Figure~\ref{fig:stage-1-3}. The results view provides
  tabs to display the source code of the implementation being tested,
  the shows, the test case source code, and the output resulting from
  the running of the test.
\item[Stage 2: Peer-Testing \& Feedback] When the teachers moves the
  platform to peer-testing stages, solutions cannot be modified and
  students are now expected to test the solutions of their peers (each
  student belongs to a group of peers). In the same way they requested
  runs of tests on their own implementation, they can now run tests on
  their peers' solutions, see Figure~\ref{fig:stage-2-1}. The result
  view now contains a discussion board where the student-tester and
  the student-developer can enter in a feedback discussion about a
  run, see Figure~\ref{fig:stage-2-2}.
\item[Stage 3: Teacher Feedback] After a deadline announced in
  advance to the students, the teacher moves the platform to the last
  stage where no new upload of solution or tests are allowed and where
  students are expected to write a reflective report of their
  peer-testing experience. The teacher will be giving feedback and
  marks on the report and the students' uploads.
\end{description}

\begin{table*}
  \centering
  \begin{tabularx}{\hsize}{clc@{}c@{}c@{}cLL}
    \toprule
    \multicolumn{2}{l}{\bf Stage}
    & \rot{Solutions upload}
    & \rot{Tests upload}
    & \rot{Self-testing}
    & \rot{Peer-testing}
    & \bf Students
    & \bf Teachers \\
    % Stage & \rot{Vertical} &  \\
    \midrule
    0 & Coursework Setup
    & \NoCheck
    & \NoCheck
    & \NoCheck
    & \NoCheck
    & Not involved at this stage.
    & Setting Coursework specification, and program
    interfaces. Enrolling students and setting up
    peer groups. Preparing oracle solution and tests. \\
    \cmidrule(rl){2-8}
    1 & Development \& Self-Testing
    & \Check
    & \Check
    & \Check
    & \NoCheck
    & Develop and upload solutions and tests. Run own tests on own
    solutions and on oracle solution. Run provided tests on own
    solutions.
    & Observing individual progress. Amending peer groups. Adding
    extra tests or solutions. \\
    \cmidrule(rl){2-8}
    2 & Peer-Testing \& Feedback
    & \NoCheck
    & \Check
    & \Check
    & \Check
    & Run tests on peers' solutions, review peers' code to send
      feedback. Receive tests and feedback. Enter discussions with peers.
    & Monitoring students discussions. Prepare individual feedback. \\
    \cmidrule(rl){2-8}
    3 & Teacher Feedback
    & \NoCheck
    & \NoCheck
    & \NoCheck
    & \NoCheck
    & Submit an experience report detailing how they would take
    on-board feedback they received to improve their program and how
    they compare their solution with their peers.
    & Provide group feedback at group level.
    \\
    \bottomrule
  \end{tabularx}
  \caption{Peer-testing stages implemented in the Web platform}
  \begin{minipage}{.8\linewidth}
    The table indicates for each stage whether students are allowed to
    submit artifacts solutions, to submit test cases, to test their
    own implementation, to view and test their peers solutions. It
    also describes the expected actions by students and teachers
    participants.
  \end{minipage}
  \label{tab:stages}
\end{table*}

\section{Related Works}
\label{sec:related-works}

In this section we explore approaches that have been explored to
provide peer feedback on programming. We compare these related work
with our platform. This review of related work is by no means
exhaustive.

\subsection{Testing in Peer Learning Activities}

\citet{goldwasser_gimmick_2002} started a peer testing initiative on a
Data Structures course.  The main objective of this was to get
students taking the course to submit, along with coursework, a test
case to run on the program with the aim of finding bugs in other
students' programs. The main objective in this case was to focus
learning on software testing, and the peer testing aspect was
secondary, however the study does hold many parallels with our
platform.
The submitted coursework solutions would be collected, along with each
test case. Then, using an automated system each of the test cases
would be run against each of the solutions. The major downside to this
cross testing activity being that there is $x^2$ complexity in terms
of the submission count.

\citet{Clark_ACE-2004} reports on a 6-year long experiment in having
teams of students test their peers' implementations in the context of
a software engineering project. The study highlights the positive
learning outcomes of such activity to increase the quality of the
work, the quality of group collaboration, and in making students
understand the necessity of testing. Our platform aims to replicate
these results in the context of individual coursework where the
management pf peer testing and peer feedback in general is an issue

\citet{ReiFinTer_Group-2009} presented the system they developed for
program peer assessment also called peer code review in a non
educational context. As part of the peer assessment, their system
required student reviewers to submit test cases results as part of
their review as well as an evaluation of the code quality. They
structured the peer assessment written comments through a proforma
where reviewers are to indicate the outcome of each test case
(choosing between \emph{does not compile}, \emph{fatal error/crash},
\emph{incorrect result(s)}, or \emph{correct result(s)}) and are to rate
the code quality following criteria (\emph{code formatting},
\emph{conventions}, \emph{documentation}, \emph{interface design},
\emph{modularity}). Similarly, feedback is provided back on the review
following criteria (\emph{accuracy}, \emph{helpfulness},
\emph{reviewer knowledge}, \emph{fairness}).
Although test cases are the cornerstone of feedback like in our
platform, these tests are not run by their system but left to the
students to manage, which makes reusing test cases on multiple
solutions more tedious to run. The categorisation of test outputs, and
feedback seem to be of benefits for structuring the reviews and
automating the grading of solutions and reviews. We would consider
this approach in future extension of our platform.

\citet{Zeller_ITiCSE-2000} describe the Praktomat system they have
developed which manages students' peer feedback on programs. The
system performs automatic testing of the student submissions. The
submission has to pass public
tests (provided by the lecturer and available to the students)  to be
submitted. The system runs additional secret tests (provided by the
lecturer but not shared with the students) for assessment by
the teacher.  Once submitted, students could review each other's
program. The students themselves are not required to submit or perform
tests. While student considered positively the fairness of testing all
submissions with the same set of tests, they initially found the
fault-finding system to be fussy.

\subsection{Online Peer Learning}

A study by \citet{davies_computerized_2000} was conducted in a
University's communications and networking module using a peer
assessment system . This peer assessment system aimed to help students
easily provide feedback on reports, in addition to attempting to
locate plagiarism. The \ac{CAP} provided an interface to let students
read the report of another student. They could then provide feedback
on the report and follow references on the report to identify any
\textit{copy \& paste} plagiarism. The study followed the results
students achieved over a sequence of 4 tasks (prepare a report, answer
multiple choice test, engage in peer marking of the reports, answer a
final multiple choice test).  The experiment looked to see if the peer
marking was at all helpful in improving the marks of students between
the two tests. The results found that the peer marking was very useful
for students who had initially performed poorly. Student comments on
the experience would suggest that they found the peer assessment
process both enjoyable and informative. There is also evidence from
this feedback indicating the importance of maintaining anonymity, as
some students felt it would be difficult to do the marking
non-anonymously.  This study also revealed that students learnt a lot
from the ``repetitive nature of the marking''.

To determine whether students can benefit from peer assessment in an
online, \textit{technology-enhanced} environment, a research project
was conducted by \citet{keppell_peer_2006}. This project
involved some university lecturers, who re-structured their courses to
involve more peer learning, with the aid of a \ac{VLE}.
The project took place over three different case studies. The
different studies used journal, discussion and file sharing features
of the \ac{VLE} to enhance two courses in fashion design, art education
and the design of a new learning website. The evidence
collected during these case studies, showed that
students found this assessment to be fairer than just against teacher;
students appreciated the ongoing peer critique performed through
reflective journals;
teachers felt the instant nature of feedback was very useful;
the support offered by the assisting technology of the \ac{VLE}
encouraged collaboration.
The authors of the project also suggest that unless the peer
assessment within the \acp{VLE} is a marked process (against a
students grade), students may be unwilling to participate.

A number of other peer assessment systems exist, see
\cite{LuxtonReilly_CSE-2009-v19n4} for a more substantial survey.

\subsection{Automatic Assessment}

In the 1990s, the University of Nottingham produced \emph{Ceilidh}, a
system that had the intent of automating and improving the assessment
process for C Language Coursework. Later improvements were made by
Heriot-Watt University adding support for \ac{SML}
\cite{foubister_automatic_1997}, the system was later renamed
CourseMarker \cite{higgins_coursemarker_2003}. This tool was used to
guide students through the coursework, and then check the correctness
of submissions. Ceilidh offered a skeleton answer to students, which
pointed out any special language features that should be used. The
submitted solutions are checked for correctness with a mixture of
verifying output against a model solution, and checking the style of
the code (e.g line count, use of certain function names, etc).  In
particular for \ac{SML}, students were asked to provide the types for
written functions, as an additional test of understanding.  To check
if such a system used for automatic assessment of students would be
useful, the results of the final exam mark were compared to those
marks from previous years where Ceilidh was not used. The study found
that there was no real improvement made by Ceilidh, but most
importantly there were no detrimental effects. Some positive side
effects were observed: the time taken for students to receive feedback
on the coursework was markedly decreased, and the tracking of progress
offered by the tool allowed teachers to more easily identify any
students that were having serious difficulties.

A number of other automated programming assessment systems and
approaches exist, see \cite{AlaMutka_CSE-2005-v15n2} for a more
complete investigation.

\subsection{Learning Game and Competition}

\citet{TildHaXieGulBis_ICSE-2013} report on the system Pex4Fun they
had designed which automated testing to build a gaming-puzzle.  In
this system students are expected to attempt to replicate the
sample-solution program hidden from them from a empty or faulty
implementation. The system uses symbolic execution of the student's
version and the secret version to provide feedback to the student.
The system was successful as a platform for teaching and learning to
program.  Pex4Fun later evolved into a new game called Code Hunt
\citep{BisHorXieTilHal_ICSE-2015,HorBisdHaTil_CHESE-2015} which
combines multiple programming puzzles of the kind Pex4Fun uses. By
organising puzzles into difficulty levels and combining these to form
a contest, Code Hunt creates an incentive to enter the
coding-game. The levels are re-evaluated using players' statistics.
Beyond being serious games, an interesting aspect of both these
systems is the absence of specification in the coding exercises which
makes the participants to reflect on testing results to
reverse-engineer the secret program.  They also limit the feedback
given to a small number of inputs and results not to overwhelm the
participant with information which is something our platform currently
leave to the peer interaction between students to manage.

\citet{Rue+Hic+Par+Lev+Mem+Pla+Mar_CCS-2016} introduce the programming
contest \emph{Build-it, Break-it, Fix-it} that uses the concept of a
peer security testing to judge the success of various programming
assignments. This success is measured both in terms of general
correctness and specifically in the context of security. Although the
contest was used in the context of a \ac{MOOC}, its aim is
specifically targeting security and therefore could not directly be
employed for educational programming peer assessment.
The contest requires participants to \textit{build} a working solution
that matches functional specifications, as well as being as secure as
possible. If the solution can be built and the functionality verified
by an automated system, then the team can move on to the next
stage. Following this, testers will attempt to \textit{break} the
security of the solution, and points will be allocated according to a
zero-sum-game. The builders can then gain points by \textit{fixing}
their solutions.  Throughout the contest, there is a central web-based
infrastructure that manages running the contest website (with
scoreboards, etc.), listening to builder git repositories, and also
running and recording test results.

\section{Conclusion}
\label{sec:conclusion}

We have presented in this paper a Web platform for peer-testing of
programming code and the process of its development which was informed
by both teachers and students. While the platform is still a work in
progress, we will be deploying it this year in our Data Structures and
Algorithms course which is simultaneously taught at Heriot-Watt
University's Edinburgh and Dubai campuses. We will be collecting
feedback on the platform from the students and expect as a result to
discover new issues and avenues for improvements.

The platform combines the delivery of coursework assignments within
\ac{CS} programming courses, the self testing of solutions to said
coursework, and subsequently the peer-testing that can take place
after submission of the coursework is complete.

\subsection{Future Works}

We are contemplating a number of options to further develop our
platform. While we have developed a system from scratch we believe
that some features of the platform would be better handled by
specialised systems. The platform would therefore benefit from a Git
integration for managing students codes as is for instance the system
by \citet{Rue+Hic+Par+Lev+Mem+Pla+Mar_CCS-2016}, from an integration
into \acp{VLE} for the management of students submission and grading,
and from an integration with existing generic peer feedback or peer
assessment systems to manage peer interactions. As discussed in
Section~\ref{sec:student-requirements}, the feedback discussion board
could be extended to include teacher interactions and forum-style
discussions. Furthermore, the platform would benefit from integrating
peer code review
as done by \citet{Trytten_SIGSE-2005} and \citet{HunAgrAgr_TOCE-2013}
and offer a classification of tests and errors as
proposed by \citep{SonMul_CSE-2012-v22n4} and
\citep{ReiFinTer_Group-2009}.

%% Acknowledgments
\begin{acks}                            %% acks environment is optional
                                        %% contents suppressed with 'anonymous'
  This work was in part supported by a Heriot-Watt University and
  \acs{QAA} joint funding.
  %% Commands \grantsponsor{<sponsorID>}{<name>}{<url>} and
  %% \grantnum[<url>]{<sponsorID>}{<number>} should be used to
  %% acknowledge financial support and will be used by metadata
  %% extraction tools.
  % This material is based upon work supported by the
  % \grantsponsor{GS100000001}{National Science
  %   Foundation}{http://dx.doi.org/10.13039/100000001} under Grant
  % No.~\grantnum{GS100000001}{nnnnnnn} and Grant
  % No.~\grantnum{GS100000001}{mmmmmmm}.  Any opinions, findings, and
  % conclusions or recommendations expressed in this material are those
  % of the author and do not necessarily reflect the views of the
  % National Science Foundation.
\end{acks}

%%% -*-BibTeX-*-
%%% Do NOT edit. File created by BibTeX with style
%%% ACM-Reference-Format-Journals [18-Jan-2012].

%% Appendix
%\appendix
%\section{Appendix}
%Text of appendix \ldots

\begin{acronym}[MOOC]
  \acro{CAP}{Computerised Assessment with Plagiarism}
  \acro{CS}{Computer Science}
  \acro{MOOC}{Massive Open Online Course}
  \acro{SML}{Standard ML}
  \acro{TDD}{Test-Driven Development}
  \acro{VLE}{Virtual Learning Environment}
  \acro{QAA}{Quality Assurance Agency for Higher Education}
  \acro{HE}{Higher Education}
\end{acronym}


\begin{thebibliography}{00}

%%% ====================================================================
%%% NOTE TO THE USER: you can override these defaults by providing
%%% customized versions of any of these macros before the \bibliography
%%% command.  Each of them MUST provide its own final punctuation,
%%% except for \shownote{}, \showDOI{}, and \showURL{}.  The latter two
%%% do not use final punctuation, in order to avoid confusing it with
%%% the Web address.
%%%
%%% To suppress output of a particular field, define its macro to expand
%%% to an empty string, or better, \unskip, like this:
%%%
%%% \newcommand{\showDOI}[1]{\unskip}   % LaTeX syntax
%%%
%%% \def \showDOI #1{\unskip}           % plain TeX syntax
%%%
%%% ====================================================================

\ifx \showCODEN    \undefined \def \showCODEN     #1{\unskip}     \fi
\ifx \showDOI      \undefined \def \showDOI       #1{#1}\fi
\ifx \showISBNx    \undefined \def \showISBNx     #1{\unskip}     \fi
\ifx \showISBNxiii \undefined \def \showISBNxiii  #1{\unskip}     \fi
\ifx \showISSN     \undefined \def \showISSN      #1{\unskip}     \fi
\ifx \showLCCN     \undefined \def \showLCCN      #1{\unskip}     \fi
\ifx \shownote     \undefined \def \shownote      #1{#1}          \fi
\ifx \showarticletitle \undefined \def \showarticletitle #1{#1}   \fi
\ifx \showURL      \undefined \def \showURL       {\relax}        \fi
% The following commands are used for tagged output and should be
% invisible to TeX
\providecommand\bibfield[2]{#2}
\providecommand\bibinfo[2]{#2}
\providecommand\natexlab[1]{#1}
\providecommand\showeprint[2][]{arXiv:#2}

\bibitem[\protect\citeauthoryear{Ala-Mutka}{Ala-Mutka}{2005}]%
        {AlaMutka_CSE-2005-v15n2}
\bibfield{author}{\bibinfo{person}{Kirsti~M. Ala-Mutka}.}
  \bibinfo{year}{2005}\natexlab{}.
\newblock \showarticletitle{A {{Survey}} of {{Automated Assessment Approaches}}
  for {{Programming Assignments}}}.
\newblock \bibinfo{journal}{{\em Computer Science Education\/}}
  \bibinfo{volume}{15}, \bibinfo{number}{2} (\bibinfo{date}{June}
  \bibinfo{year}{2005}), \bibinfo{pages}{83--102}.
\newblock
\showISSN{0899-3408}
\showDOI{%
\url{https://doi.org/10.1080/08993400500150747}}


\bibitem[\protect\citeauthoryear{Bishop, Horspool, Xie, Tillmann, and
  De~Halleux}{Bishop et~al\mbox{.}}{2015}]%
        {BisHorXieTilHal_ICSE-2015}
\bibfield{author}{\bibinfo{person}{Judith Bishop}, \bibinfo{person}{R.~Nigel
  Horspool}, \bibinfo{person}{Tao Xie}, \bibinfo{person}{Nikolai Tillmann},
  {and} \bibinfo{person}{Jonathan De~Halleux}.}
  \bibinfo{year}{2015}\natexlab{}.
\newblock \showarticletitle{Code {{Hunt}}: {{Experience}} with {{Coding
  Contests}} at {{Scale}}}. In \bibinfo{booktitle}{{\em 2015 {{IEEE}}/{{ACM}}
  37th {{IEEE International Conference}} on {{Software Engineering}}}},
  Vol.~\bibinfo{volume}{2}. \bibinfo{pages}{398--407}.
\newblock
\showDOI{%
\url{https://doi.org/10.1109/ICSE.2015.172}}


\bibitem[\protect\citeauthoryear{Clark}{Clark}{2004}]%
        {Clark_ACE-2004}
\bibfield{author}{\bibinfo{person}{Nicole Clark}.}
  \bibinfo{year}{2004}\natexlab{}.
\newblock \showarticletitle{Peer {{Testing}} in {{Software Engineering
  Projects}}}. In \bibinfo{booktitle}{{\em Proceedings of the {{Sixth
  Australasian Conference}} on {{Computing Education}} - {{Volume}} 30}} {\em
  (\bibinfo{series}{ACE '04})}. \bibinfo{publisher}{{Australian Computer
  Society, Inc.}}, \bibinfo{address}{Darlinghurst, Australia, Australia},
  \bibinfo{pages}{41--48}.
\newblock
\showURL{%
\url{http://dl.acm.org/citation.cfm?id=979968.979974}}


\bibitem[\protect\citeauthoryear{Davies}{Davies}{2000}]%
        {davies_computerized_2000}
\bibfield{author}{\bibinfo{person}{Phil Davies}.}
  \bibinfo{year}{2000}\natexlab{}.
\newblock \showarticletitle{Computerized Peer Assessment}.
\newblock \bibinfo{journal}{{\em Innovations in Education \& Training
  International\/}} \bibinfo{volume}{37}, \bibinfo{number}{4}
  (\bibinfo{year}{2000}), \bibinfo{pages}{346--355}.
\newblock
\showISSN{1355-8005, 1469-8420}
\showDOI{%
\url{https://doi.org/10.1080/135580000750052955}}


\bibitem[\protect\citeauthoryear{Falchikov}{Falchikov}{2013}]%
        {falchikov_improving_2013}
\bibfield{author}{\bibinfo{person}{Nancy Falchikov}.}
  \bibinfo{year}{2013}\natexlab{}.
\newblock \bibinfo{booktitle}{{\em Improving Assessment Through Student
  Involvement: Practical Solutions for Aiding Learning in Higher and Further
  Education}}.
\newblock \bibinfo{publisher}{Routledge}.
\newblock
\showISBNx{978-1-134-39575-0}


\bibitem[\protect\citeauthoryear{Foubister, Michaelson, and Tomes}{Foubister
  et~al\mbox{.}}{1997}]%
        {foubister_automatic_1997}
\bibfield{author}{\bibinfo{person}{Sandra~P. Foubister}, \bibinfo{person}{G.~J.
  Michaelson}, {and} \bibinfo{person}{Nils Tomes}.}
  \bibinfo{year}{1997}\natexlab{}.
\newblock \showarticletitle{Automatic assessment of elementary Standard {ML}
  programs using Ceilidh}.
\newblock \bibinfo{journal}{{\em Journal of Computer Assisted Learning\/}}
  \bibinfo{volume}{13}, \bibinfo{number}{2} (\bibinfo{year}{1997}),
  \bibinfo{pages}{99--108}.
\newblock
\showDOI{%
\url{https://doi.org/10.1046/j.1365-2729.1997.00012.x}}


\bibitem[\protect\citeauthoryear{Goldwasser}{Goldwasser}{2002}]%
        {goldwasser_gimmick_2002}
\bibfield{author}{\bibinfo{person}{Michael~H. Goldwasser}.}
  \bibinfo{year}{2002}\natexlab{}.
\newblock \showarticletitle{A Gimmick to Integrate Software Testing Throughout
  the Curriculum}. In \bibinfo{booktitle}{{\em 33rd {SIGCSE} Technical
  Symposium on Computer Science Education}} {\em (\bibinfo{series}{{SIGCSE}
  '02})}. \bibinfo{publisher}{{ACM}}, \bibinfo{pages}{271--275}.
\newblock
\showISBNx{978-1-58113-473-5}
\showDOI{%
\url{https://doi.org/10.1145/563340.563446}}


\bibitem[\protect\citeauthoryear{Grov, Hamdan, Kumar, Maarek, McGregor, Shaikh,
  Wells, and Zantout}{Grov et~al\mbox{.}}{2017}]%
        {GroHamKumMaaMcGrShaWelZan_STEM-HE-2017}
\bibfield{author}{\bibinfo{person}{Gudmund Grov}, \bibinfo{person}{Mohammad
  Hamdan}, \bibinfo{person}{Smitha~S Kumar}, \bibinfo{person}{Manuel Maarek},
  \bibinfo{person}{Léon McGregor}, \bibinfo{person}{Talal Shaikh},
  \bibinfo{person}{J.B. Wells}, {and} \bibinfo{person}{Hind Zantout}.}
  \bibinfo{year}{2017}\natexlab{}.
\newblock \showarticletitle{Transition from passive learner to critical
  evaluator through peer-testing of programming artifacts}. In
  \bibinfo{booktitle}{{\em Horizons in STEM Higher Education Conference: Making
  Connections and Sharing Pedagogy}}. \bibinfo{address}{Edinburgh, UK}.
\newblock
\newblock
\shownote{(presentation in June 2017, paper in preparation).}


\bibitem[\protect\citeauthoryear{Higgins, Hegazy, Symeonidis, and
  Tsintsifas}{Higgins et~al\mbox{.}}{2003}]%
        {higgins_coursemarker_2003}
\bibfield{author}{\bibinfo{person}{Colin Higgins}, \bibinfo{person}{Tarek
  Hegazy}, \bibinfo{person}{Pavlos Symeonidis}, {and}
  \bibinfo{person}{Athanasios Tsintsifas}.} \bibinfo{year}{2003}\natexlab{}.
\newblock \showarticletitle{The {CourseMarker} {CBA} System: Improvements over
  Ceilidh}.
\newblock \bibinfo{journal}{{\em Education and Information Technologies\/}}
  \bibinfo{volume}{8}, \bibinfo{number}{3} (\bibinfo{year}{2003}),
  \bibinfo{pages}{287--304}.
\newblock
\showISSN{1360-2357, 1573-7608}
\showDOI{%
\url{https://doi.org/10.1023/A:1026364126982}}


\bibitem[\protect\citeauthoryear{Horspool, Bishop, {de Halleux}, and
  Tillmann}{Horspool et~al\mbox{.}}{2015}]%
        {HorBisdHaTil_CHESE-2015}
\bibfield{author}{\bibinfo{person}{R.~Nigel Horspool}, \bibinfo{person}{Judith
  Bishop}, \bibinfo{person}{Jonathan {de Halleux}}, {and}
  \bibinfo{person}{Nikolai Tillmann}.} \bibinfo{year}{2015}\natexlab{}.
\newblock \showarticletitle{Experience with {{Constructing Code Hunt
  Contests}}}. In \bibinfo{booktitle}{{\em Proceedings of the 1st
  {{International Workshop}} on {{Code Hunt Workshop}} on {{Educational
  Software Engineering}}}} {\em (\bibinfo{series}{CHESE 2015})}.
  \bibinfo{publisher}{{ACM}}, \bibinfo{address}{New York, NY, USA},
  \bibinfo{pages}{1--4}.
\newblock
\showISBNx{978-1-4503-3711-3}
\showDOI{%
\url{https://doi.org/10.1145/2792404.2792405}}


\bibitem[\protect\citeauthoryear{Hundhausen, Agrawal, and Agarwal}{Hundhausen
  et~al\mbox{.}}{2013}]%
        {HunAgrAgr_TOCE-2013}
\bibfield{author}{\bibinfo{person}{Christopher~D. Hundhausen},
  \bibinfo{person}{Anukrati Agrawal}, {and} \bibinfo{person}{Pawan Agarwal}.}
  \bibinfo{year}{2013}\natexlab{}.
\newblock \showarticletitle{Talking {{About Code}}: {{Integrating Pedagogical
  Code Reviews}} into {{Early Computing Courses}}}.
\newblock \bibinfo{journal}{{\em Trans. Comput. Educ.\/}} \bibinfo{volume}{13},
  \bibinfo{number}{3} (\bibinfo{date}{Aug.} \bibinfo{year}{2013}),
  \bibinfo{pages}{14:1--14:28}.
\newblock
\showISSN{1946-6226}
\showDOI{%
\url{https://doi.org/10.1145/2499947.2499951}}


\bibitem[\protect\citeauthoryear{Keppell, Au, Ma, and Chan}{Keppell
  et~al\mbox{.}}{2006}]%
        {keppell_peer_2006}
\bibfield{author}{\bibinfo{person}{Mike Keppell}, \bibinfo{person}{Eliza Au},
  \bibinfo{person}{Ada Ma}, {and} \bibinfo{person}{Christine Chan}.}
  \bibinfo{year}{2006}\natexlab{}.
\newblock \showarticletitle{Peer learning and learning-oriented assessment in
  technology-enhanced environments}.
\newblock \bibinfo{journal}{{\em Assessment \& Evaluation in Higher
  Education\/}} \bibinfo{volume}{31}, \bibinfo{number}{4} (\bibinfo{date}{Aug.}
  \bibinfo{year}{2006}), \bibinfo{pages}{453--464}.
\newblock
\showISSN{0260-2938, 1469-297X}
\showDOI{%
\url{https://doi.org/10.1080/02602930600679159}}


\bibitem[\protect\citeauthoryear{Laboon}{Laboon}{2016}]%
        {laboon_friendly_2016}
\bibfield{author}{\bibinfo{person}{Bill Laboon}.}
  \bibinfo{year}{2016}\natexlab{}.
\newblock \bibinfo{booktitle}{{\em A Friendly Introduction to Software
  Testing\/} (\bibinfo{edition}{1} ed.)}.
\newblock \bibinfo{publisher}{CreateSpace Independent Publishing Platform}.
\newblock


\bibitem[\protect\citeauthoryear{Li}{Li}{2016}]%
        {li_role_2016}
\bibfield{author}{\bibinfo{person}{Lan Li}.} \bibinfo{year}{2016}\natexlab{}.
\newblock \showarticletitle{The role of anonymity in peer assessment}.
\newblock \bibinfo{journal}{{\em Assessment \& Evaluation in Higher
  Education\/}} (\bibinfo{date}{April} \bibinfo{year}{2016}),
  \bibinfo{pages}{1--12}.
\newblock
\showISSN{0260-2938, 1469-297X}
\showDOI{%
\url{https://doi.org/10.1080/02602938.2016.1174766}}


\bibitem[\protect\citeauthoryear{Li, Liu, and Steckelberg}{Li
  et~al\mbox{.}}{2010}]%
        {li_assessor_2010}
\bibfield{author}{\bibinfo{person}{Lan Li}, \bibinfo{person}{Xiongyi Liu},
  {and} \bibinfo{person}{Allen~L. Steckelberg}.}
  \bibinfo{year}{2010}\natexlab{}.
\newblock \showarticletitle{Assessor or assessee: How student learning improves
  by giving and receiving peer feedback}.
\newblock \bibinfo{journal}{{\em British Journal of Educational Technology\/}}
  \bibinfo{volume}{41}, \bibinfo{number}{3} (\bibinfo{date}{May}
  \bibinfo{year}{2010}), \bibinfo{pages}{525--536}.
\newblock
\showISSN{00071013, 14678535}
\showDOI{%
\url{https://doi.org/10.1111/j.1467-8535.2009.00968.x}}


\bibitem[\protect\citeauthoryear{Luxton-Reilly}{Luxton-Reilly}{2009}]%
        {LuxtonReilly_CSE-2009-v19n4}
\bibfield{author}{\bibinfo{person}{Andrew Luxton-Reilly}.}
  \bibinfo{year}{2009}\natexlab{}.
\newblock \showarticletitle{A Systematic Review of Tools That Support Peer
  Assessment}.
\newblock \bibinfo{journal}{{\em Computer Science Education\/}}
  \bibinfo{volume}{19}, \bibinfo{number}{4} (\bibinfo{date}{Dec.}
  \bibinfo{year}{2009}), \bibinfo{pages}{209--232}.
\newblock
\showISSN{0899-3408}
\showDOI{%
\url{https://doi.org/10.1080/08993400903384844}}


\bibitem[\protect\citeauthoryear{McGregor}{McGregor}{2017}]%
        {McGregor_BSc-2017}
\bibfield{author}{\bibinfo{person}{Léon McGregor}.}
  \bibinfo{year}{2017}\natexlab{}.
\newblock \bibinfo{title}{Web Platform for Code Peer-Testing}.
\newblock \bibinfo{howpublished}{BSc Honours dissertation, Heriot-Watt
  University}.   (\bibinfo{date}{April} \bibinfo{year}{2017}).
\newblock


\bibitem[\protect\citeauthoryear{Reily, Finnerty, and Terveen}{Reily
  et~al\mbox{.}}{2009}]%
        {ReiFinTer_Group-2009}
\bibfield{author}{\bibinfo{person}{Ken Reily}, \bibinfo{person}{Pam~Ludford
  Finnerty}, {and} \bibinfo{person}{Loren Terveen}.}
  \bibinfo{year}{2009}\natexlab{}.
\newblock \showarticletitle{Two {{Peers Are Better Than One}}: {{Aggregating
  Peer Reviews}} for {{Computing Assignments}} Is {{Surprisingly Accurate}}}.
  In \bibinfo{booktitle}{{\em Proceedings of the {{ACM}} 2009 {{International
  Conference}} on {{Supporting Group Work}}}} {\em (\bibinfo{series}{GROUP
  '09})}. \bibinfo{publisher}{{ACM}}, \bibinfo{address}{New York, NY, USA},
  \bibinfo{pages}{115--124}.
\newblock
\showISBNx{978-1-60558-500-0}
\showDOI{%
\url{https://doi.org/10.1145/1531674.1531692}}


\bibitem[\protect\citeauthoryear{Ruef, Hicks, Parker, Levin, Mazurek, and
  Mardziel}{Ruef et~al\mbox{.}}{2016}]%
        {Rue+Hic+Par+Lev+Mem+Pla+Mar_CCS-2016}
\bibfield{author}{\bibinfo{person}{Andrew Ruef}, \bibinfo{person}{Michael~W.
  Hicks}, \bibinfo{person}{James Parker}, \bibinfo{person}{Dave Levin},
  \bibinfo{person}{Michelle~L. Mazurek}, {and} \bibinfo{person}{Piotr
  Mardziel}.} \bibinfo{year}{2016}\natexlab{}.
\newblock \showarticletitle{Build It, Break It, Fix It: Contesting Secure
  Development}. In \bibinfo{booktitle}{{\em {ACM} {SIGSAC} Conference on
  Computer and Communications Security (CCS)}}. \bibinfo{address}{Vienna,
  Austria}, \bibinfo{pages}{690--703}.
\newblock
\showDOI{%
\url{https://doi.org/10.1145/2976749.2978382}}


\bibitem[\protect\citeauthoryear{Sadler and Good}{Sadler and Good}{2006}]%
        {sadler_impact_2006}
\bibfield{author}{\bibinfo{person}{Philip~M. Sadler} {and}
  \bibinfo{person}{Eddie Good}.} \bibinfo{year}{2006}\natexlab{}.
\newblock \showarticletitle{The Impact of Self- and Peer-Grading on Student
  Learning}.
\newblock \bibinfo{journal}{{\em Educational Assessment\/}}
  \bibinfo{volume}{11}, \bibinfo{number}{1} (\bibinfo{year}{2006}),
  \bibinfo{pages}{1--31}.
\newblock
\showISSN{1062-7197}
\showDOI{%
\url{https://doi.org/10.1207/s15326977ea1101_1}}


\bibitem[\protect\citeauthoryear{Sitthiworachart and Joy}{Sitthiworachart and
  Joy}{2004}]%
        {sitthiworachart_effective_2004}
\bibfield{author}{\bibinfo{person}{Jirarat Sitthiworachart} {and}
  \bibinfo{person}{Mike Joy}.} \bibinfo{year}{2004}\natexlab{}.
\newblock \showarticletitle{Effective peer assessment for learning computer
  programming}. \bibinfo{publisher}{{ACM} Press}, \bibinfo{pages}{122}.
\newblock
\showISBNx{978-1-58113-836-8}
\showDOI{%
\url{https://doi.org/10.1145/1007996.1008030}}


\bibitem[\protect\citeauthoryear{Smith, Tessler, Kramer, and Lin}{Smith
  et~al\mbox{.}}{2012}]%
        {SmiTesKraLin_ICER-2012}
\bibfield{author}{\bibinfo{person}{Joanna Smith}, \bibinfo{person}{Joe
  Tessler}, \bibinfo{person}{Elliot Kramer}, {and} \bibinfo{person}{Calvin
  Lin}.} \bibinfo{year}{2012}\natexlab{}.
\newblock \showarticletitle{Using {{Peer Review}} to {{Teach Software
  Testing}}}. In \bibinfo{booktitle}{{\em Proceedings of the {{Ninth Annual
  International Conference}} on {{International Computing Education
  Research}}}} {\em (\bibinfo{series}{ICER '12})}. \bibinfo{publisher}{{ACM}},
  \bibinfo{address}{New York, NY, USA}, \bibinfo{pages}{93--98}.
\newblock
\showISBNx{978-1-4503-1604-0}
\showDOI{%
\url{https://doi.org/10.1145/2361276.2361295}}


\bibitem[\protect\citeauthoryear{S{\o}ndergaard and Mulder}{S{\o}ndergaard and
  Mulder}{2012}]%
        {SonMul_CSE-2012-v22n4}
\bibfield{author}{\bibinfo{person}{Harald S{\o}ndergaard} {and}
  \bibinfo{person}{Raoul~A. Mulder}.} \bibinfo{year}{2012}\natexlab{}.
\newblock \showarticletitle{Collaborative Learning through Formative Peer
  Review: Pedagogy, Programs and Potential}.
\newblock \bibinfo{journal}{{\em Computer Science Education\/}}
  \bibinfo{volume}{22}, \bibinfo{number}{4} (\bibinfo{date}{Dec.}
  \bibinfo{year}{2012}), \bibinfo{pages}{343--367}.
\newblock
\showISSN{0899-3408}
\showDOI{%
\url{https://doi.org/10.1080/08993408.2012.728041}}


\bibitem[\protect\citeauthoryear{Tillmann, {de Halleux}, Xie, Gulwani, and
  Bishop}{Tillmann et~al\mbox{.}}{2013}]%
        {TildHaXieGulBis_ICSE-2013}
\bibfield{author}{\bibinfo{person}{Nikolai Tillmann}, \bibinfo{person}{Jonathan
  {de Halleux}}, \bibinfo{person}{Tao Xie}, \bibinfo{person}{Sumit Gulwani},
  {and} \bibinfo{person}{Judith Bishop}.} \bibinfo{year}{2013}\natexlab{}.
\newblock \showarticletitle{Teaching and {{Learning Programming}} and
  {{Software Engineering}} via {{Interactive Gaming}}}. In
  \bibinfo{booktitle}{{\em Proceedings of the 2013 {{International Conference}}
  on {{Software Engineering}}}} {\em (\bibinfo{series}{ICSE '13})}.
  \bibinfo{publisher}{{IEEE Press}}, \bibinfo{address}{Piscataway, NJ, USA},
  \bibinfo{pages}{1117--1126}.
\newblock
\showISBNx{978-1-4673-3076-3}
\showDOI{%
\url{https://doi.org/10.1109/ICSE.2013.6606662}}


\bibitem[\protect\citeauthoryear{Topping}{Topping}{2009}]%
        {topping_peer_2009}
\bibfield{author}{\bibinfo{person}{Keith~J. Topping}.}
  \bibinfo{year}{2009}\natexlab{}.
\newblock \showarticletitle{Peer Assessment}.
\newblock \bibinfo{journal}{{\em Theory Into Practice\/}} \bibinfo{volume}{48},
  \bibinfo{number}{1} (\bibinfo{year}{2009}), \bibinfo{pages}{20--27}.
\newblock
\showISSN{00405841}
\showDOI{%
\url{https://doi.org/10.1080/00405840802577569}}


\bibitem[\protect\citeauthoryear{Trytten}{Trytten}{2005}]%
        {Trytten_SIGSE-2005}
\bibfield{author}{\bibinfo{person}{Deborah~A. Trytten}.}
  \bibinfo{year}{2005}\natexlab{}.
\newblock \showarticletitle{A {{Design}} for {{Team Peer Code Review}}}. In
  \bibinfo{booktitle}{{\em Proceedings of the 36th {{SIGCSE Technical
  Symposium}} on {{Computer Science Education}}}} {\em (\bibinfo{series}{SIGCSE
  '05})}. \bibinfo{publisher}{{ACM}}, \bibinfo{address}{New York, NY, USA},
  \bibinfo{pages}{455--459}.
\newblock
\showISBNx{978-1-58113-997-6}
\showDOI{%
\url{https://doi.org/10.1145/1047344.1047492}}


\bibitem[\protect\citeauthoryear{Zeller}{Zeller}{2000}]%
        {Zeller_ITiCSE-2000}
\bibfield{author}{\bibinfo{person}{Andreas Zeller}.}
  \bibinfo{year}{2000}\natexlab{}.
\newblock \showarticletitle{Making {{Students Read}} and {{Review Code}}}. In
  \bibinfo{booktitle}{{\em Proceedings of the 5th {{Annual SIGCSE}}/{{SIGCUE
  ITiCSEconference}} on {{Innovation}} and {{Technology}} in {{Computer Science
  Education}}}} {\em (\bibinfo{series}{ITiCSE '00})}.
  \bibinfo{publisher}{{ACM}}, \bibinfo{address}{New York, NY, USA},
  \bibinfo{pages}{89--92}.
\newblock
\showISBNx{978-1-58113-207-6}
\showDOI{%
\url{https://doi.org/10.1145/343048.343090}}


\end{thebibliography}
\end{document}